\documentclass{aa}  

\usepackage{graphicx}
\usepackage{txfonts}
\usepackage{siunitx}
\usepackage{ulem}
\usepackage{subfigure}
\usepackage{mathtools}
\usepackage{enumitem}
\usepackage[utf8]{inputenc}
\usepackage{tikz,xcolor,hyperref}

\usepackage{color}

\definecolor{red}{rgb}{1.0,0.0,0.0}
\definecolor{darkgray}{rgb}{0.66, 0.66, 0.66}
\definecolor{gray(x11gray)}{rgb}{0.75, 0.75, 0.75}
\definecolor{magenta}{rgb}{1.0, 0.0, 1.0}

\newcommand{\grey}[1]{{\color{darkgray}{#1}}}

\definecolor{lime}{HTML}{A6CE39}
\DeclareRobustCommand{\orcidicon}{%
    \begin{tikzpicture}
    \draw[lime, fill=lime] (0,0) 
    circle [radius=0.16] 
    node[white] {{\fontfamily{qag}\selectfont \tiny ID}};
    \draw[white, fill=white] (-0.0625,0.095) 
    circle [radius=0.007];
    \end{tikzpicture}
    \hspace{-2mm}
}

\newcommand{\orcidEvelyn}{\href{https://orcid.org/0000-0002-2368-6469}{\orcidicon}}
\newcommand{\orcidGaspar}{\href{https://orcid.org/0000-0002-8835-0739}{\orcidicon}}
\newcommand{\orcidMatias}{\href{https://orcid.org/0000-0003-2139-0944}{\orcidicon}}
\newcommand{\orcidPhilippe}{\href{https://orcid.org/0000-0001-5657-4837}{\orcidicon}}
\newcommand{\orcidSamuel}{\href{https://orcid.org/0000-0002-9091-2366}{\orcidicon}}
\newcommand{\orcidPaul}{\href{https://orcid.org/0000-0001-8654-0101}{\orcidicon}}
\newcommand{\orcidBenoit}{\href{https://orcid.org/0000-0002-2470-5756}{\orcidicon}}
\newcommand{\orcidJunais}{\href{https://orcid.org/0000-0002-7016-4532}{\orcidicon}}
\newcommand{\orcidYasna}{\href{https://orcid.org/0000-0001-7966-7606}{\orcidicon}}
\newcommand{\orcidThomas}{\href{https://orcid.org/0000-0003-0350-7061}{\orcidicon}}
\newcommand{\orcidPeter}{\href{https://orcid.org/0000-0003-4766-902X}{\orcidicon}}

\begin{document}

   \title{A MUSE View of the Core of the Giant Low Surface Brightness Galaxy Malin~1}


   \author{Evelyn J. Johnston\inst{1}\fnmsep\orcidEvelyn,
          Gaspar Galaz\inst{2}\orcidGaspar,
          Matias Bla\~na\inst{3}\orcidMatias,
          Philippe Amram\inst{4}\orcidPhilippe,
          Samuel Boissier\inst{4}\orcidSamuel,
          Paul Eigenthaler\inst{2}\orcidPaul,
          Beno\^it Epinat\inst{4,5}\orcidBenoit,
          Junais\inst{6}\orcidJunais,
          Yasna Ordenes-Brice\~no\inst{2}\orcidYasna,
          Thomas Puzia\inst{2}\orcidThomas
          \and
	 Peter M. Weilbacher\inst{6}\orcidPeter.
          }

   \institute{Instituto de Estudios Astrof\'isicos, Facultad de Ingenier\'ia y Ciencias, Universidad Diego Portales, Av. Ej\'ercito Libertador 441, Santiago, Chile. \\
              \email{evelyn.johnston@mail.udp.cl}
         \and
             Instituto de Astrof\'isica, Pontificia Universidad Cat\'olica de Chile, Av.~Vicu\~na Mackenna 4860, 7820436 Macul, Santiago, Chile
         \and
             Instituto de Alta Investigaci\'on, Universidad de Tarapac\'a, Casilla 7D, Arica, Chile
         \and
             Aix Marseille Univ, CNRS, CNES, LAM, Marseille, France
         \and
             Canada-France-Hawaii Telescope, 65-1238 Mamalahoa Highway, Kamuela, HI 96743, USA
         \and
             National Centre for Nuclear Research, Pasteura 7, PL-02-093 Warsaw, Poland
         \and
             Leibniz-Institut f\"ur Astrophysik Potsdam (AIP), An der Sternwarte 16, 14482 Potsdam
             }


\titlerunning{A MUSE View of the Core of Malin 1}
\authorrunning{Johnston et al}

 
  \abstract
   {}
   {The central region of the Giant Low Surface Brightness galaxy Malin~1 has long been known to have a complex morphology with evidence of a bulge, disc, and potentially a bar hosting asymmetric star formation. In this work, we use VLT/MUSE data to resolve the central region of Malin~1 in order to determine its structure.}
   {We use careful light profile fitting in every image slice of the datacube to create wavelength-dependent models of each morphological component, from which we could  cleanly extract their spectra. We then used the kinematics and emission line properties from these spectra to better understand the nature of each component extracted from our model fit.}
   {We  report the detection of a pair of distinct  sources at the centre of this galaxy with a separation of $\sim1.05\arcsec$, which corresponds to a separation on sky of $\sim1.9$~kpc. The radial velocity data of each object confirms that they both lie in the kinematic core of the galaxy, and analysis of the emission lines reveals that the central compact source is more consistent with being ionized by star formation and/or a LINER, while the off-centre compact source lies closer to the separation between star-forming galaxies and AGN. }
   {This evidence suggests that the centre of Malin~1 hosts either a bar with asymmetric star formation or two distinct components in which the off-centre compact source could either be a star-forming clump containing one or more star clusters that is in the process of falling into the core of the galaxy and which will eventually merge with the central nuclear star cluster, or a clump of gas infalling into the centre of the galaxy from either outside or from the disc and triggering star formation there.}


   \maketitle
%

\section{Introduction}

Since its discovery \citep{Bothun_1987}, Malin~1 has always been an unusual member of the class of Giant Low Surface-Brightness (GLSB) galaxies. Not only has it a very faint disk with a surface brightness of $\sim$ 28 mag arcsec$^{-2}$ in the $B$-band and an extraordinary size of $\sim 200$ kpc \citep{Moore_2006, Galaz_2015, Boissier_2016}, making it perhaps the largest single spiral observed to date, but also it has perplexing features, among them, a large reservoir of $\sim6 \times 10^{10}$ M$_\odot$ of HI \citep{pickering_1997}, and to date no direct evidence of molecular gas \citep{braine2000, das_2010, galaz2022}.
On the other hand, the inner region, of radius of $\sim10$~kpc, resembles a typical SB0/a galaxy \citep{Barth_2007}. This inner region is also complex, with evidence of a bulge+disc+bar morphology \citep{Barth_2007, Junais_2020, Saha_2021} and emission line properties indicating the presence of both star formation \citep{Junais_2020, Saha_2021} and LINER activity \citep{Barth_2007, Subramanian_2016, Junais_2020}. 

A recent study of Malin~1 by \citet{Saha_2021} has also revealed that when zoomed in, the central bar structure can be resolved into two clumps that ``give an impression of a bar-like structure'' in FUV images of the galaxy, although their Figure~5 shows that the FUV emission appears to be misaligned with respect to the bar. From their observations in the UV and of strong emission lines, they conclude that this region has undergone recent star formation ($<$ 10~Myr), and hosts hot blue stars, despite their measurements of the H$\alpha$  and FUV star-formation rates within the central $3\arcsec$ being low enough that the galaxy would normally be considered quenched. If these two clumps at the core of Malin~1 are confirmed both photometrically and kinematically to be distinct, they may reflect a rare and challenging example of a the growth of an NSC through the merging and accretion of infalling gas or star clusters.

NSCs are a common characteristic of galaxies of all masses and morphologies, with a recent study of spiral galaxies by \citet{Ashok_2023} finding 80\% of the sample hosted NSCs, but their formation and mass assembly processes are still uncertain. Two main scenarios have been proposed to explain the formation of NSCs. The first process is the \textit{in situ} scenario, in which the NSC is formed in the central few parsecs of the galaxy through star formation fuelled by infalling gas \citep{Milosavljevic_2004, Bekki_2006, Bekki_2007}. In this scenario, the star formation timescale and occurrences of later star formation are dependent on the gas reservoirs within the galaxy. The second scenario is the \textit{migration} scenario, in which a globular cluster (GC) forms  within the galaxy and migrates to the centre of the galaxy via dynamical friction \citep{Tremaine_1975}. NSCs created in this way can build up their mass at later times through mergers with other infalling GCs \citep{Andersen_2008}.  As dynamical friction is most efficient between systems with similar velocity dispersion, this process is thought to be most active in dwarf galaxies \citep[e.g.][]{Lotz_2001, Miller_2007, Neumayer_2020, Fahrion_2021}.

In reality however, it is unlikely that one single process dominates the formation of NSCs. Consequently, a hybrid scenario has been suggested- the \textit{wet-migration} scenario \citep{Guillard_2016}. In this case, a GC forms with its own gas reservoir somewhere in the disc of the galaxy. As the GC migrates to the centre of the galaxy through dynamical friction and interactions with other structures, the gas reservoir feeds the ongoing star formation, leading to the GC growing in mass as it migrates and settles at the core of the galaxy as the NSC. Through this mechanism the NSC can continue to build up its mass over time through the accretion of other infalling gas-rich GCs. 

Furthermore, it has been shown that the formation and mass assembly mechanisms for NSCs is dependent on the host galaxy mass. For example, \citet{Ordenes_2018b} found evidence for a two-regime $M_{NSC}-M_{galaxy}$ scaling relation with two very different  slopes for dwarf and giant galaxies, showing that the buildup of NSCs in dwarf galaxies uses a different mix of mechanisms than in giant galaxies. Additionally,  \citet{Turner_2012} and \citet{denBrok_2014} found evidence that mergers with migrating clusters are responsible for the growth of the NSC in low-mass galaxies, with in-situ star formation playing an important secondary role. They also found that in higher-mass galaxies, such as Malin~1 which has a stellar mass of $8.9 \times 10^{11}$~M$_\odot$ \citep{Saha_2021}, gas accretion from mergers becomes increasingly important and contributes to the in-situ star formation.

NSC mass growth through mergers with infalling GCs is thought to play a significant role in galaxies of all masses. Evidence for this scenario could come from observations of pairs of clusters at the centre of a galaxy. One such example is NGC~4654, a spiral galaxy in the Virgo Cluster, in which both nuclear star clusters show different ages of their stellar populations and whose orbits appear to be unstable, indicating that they will likely merge in the next 0.5~Gyr  \citep{Georgiev_2014}. The elliptical galaxy IC~225 has also been found to be double-nucleated, where both clusters were spectroscopically confirmed to be gravitationally bound and that the off-centre cluster is a metal-rich H{\textsc{ii}} region \citep{Gu_2006, Miller_2008}. \citet{Johnston_2020} also detected a potential pair of star clusters at the centre of the dwarf galaxy FCC~222 in the Fornax Cluster, although based on the kinematics measurements and separation, they were unable to confirm whether the infalling cluster was gravitationally bound to the NSC or was simply a foreground star cluster along the same line of sight. 

Another potential scenario that has been proposed outlines how an NSC can form or accrete mass through gas accretion \citep{Silk_1987} or a gas rich merger with another galaxy \citep{Mihos_1994}. Simulations have already shown that a galaxy with similar characteristics of Malin~1 could be formed through the encounter of three galaxies \citep{zhu2018}, and Malin~1 is already known to have a neighbour galaxy at the same redshift towards the North-East at a distance of 358~kpc that is linked by a faint stellar bridge \citep{Galaz_2015,Saha_2021}. Furthermore, \citet{Junais_2023} showed that Malin~1 has active star forming regions in the external disk, which are associated to satellites Malin~1A and Malin~1B and thus could indicate a recent global merger event that contributed to  gas accretion.

Since both high spatial-resolution imaging and spectra are necessary to detect and confirm the presence of a  star cluster accreting into an NSC, combined with the short merger timescales making them short-lived, galaxies hosting such a process are a rare phenomenon. Consequently, if the two clumps detected at the centre of Malin~1 by \citet{Saha_2021} are confirmed to be two star clusters, i.e. an NSC and an infalling star cluster, this system will provide valuable information as to how these structures assemble their mass.

However, the clumpy structure at the centre of Malin~1 may not be a bar or a GC falling into an NSC. There are several cases where other structures in the core of a galaxy can mimic these components. For example, models by  \citet{Tremaine_1995} showed that as an infalling GC migrates closer to the centre of the galaxy, its orbit becomes more unstable, until eventually it is disrupted and forms an eccentric stellar disc in the core of the galaxy. If a black hole resides at the centre of the galaxy, this disc may eventually become the accretion disc that fuels the black hole. Nuclear stellar discs that are thought to have been created from infalling GCs have been found in HCG~90-DW4 \citep{Ordenes_2016} and FCC~47 \citep{Fahrion_2019}. The models of \citet{Tremaine_1995} went further to show that apparent pairs of NSCs could really be explained as eccentric nuclear stellar discs, and this theory has been used to explain the core regions of M31 \citet{Kormendy_1999}, NGC~4486B \citep{Lauer_1996} and VCC~128 \citep{Debattista_2006}. In these cases, the nuclear disc takes on the appearance of a pair of bright sources due to the eccentricity of the orbits, where the centre of mass of the system lies close to the fainter source and that the bright source actually marked the apoapsis, the point in the orbit furthest from the centre of mass, which is brighter because the stars linger near this point in their orbits. The presence of a nuclear stellar disc can be confirmed through spectral signatures such as slow clear rotation and an asymmetric velocity dispersion profile within the nuclear disc.

This paper  presents an investigation of the nature of the core of Malin~1, using the combined spectroscopic and photometric data from the Multi-Unit Spectroscopic Explorer \citep[MUSE,][]{Bacon_2010} on the VLT to investigate the nature of the structures at the centre of this galaxy. We define the core of the galaxy as being the region at the centre of the galaxy with a radius of $\sim2$~kpc, which corresponds to region containing an extended and asymmetric emission line component and the bar structure identified by \citet{Saha_2021}. The paper is laid out as follows: Section~\ref{sec:DR} describes the observations and data reduction; Section~\ref{sec:Buddi} outlines the light profile fitting method used to model and isolate the light from each component present within the galaxy; Section~\ref{sec:analysis} presents the analysis of the data, and finally Section~\ref{sec:Conclusions} outlines our conclusions. Throughout this paper we assume a Hubble constant of $H_o = 70$~km~s$^{-1}$~Mpc$^{-1}$ \citep{Lelli_2010}, which corresponds to a projected angular scale of 1.83~kpc~arcsec$^{-1}$ and a distance of 377~Mpc  based on the line-of-sight velocity of the galaxy described in Section~\ref{sec:analysis}.

\section{Data and reduction}\label{sec:DR}
MUSE is an optical integral-field spectrograph with a field of view of $1\arcmin\!\times\!1\arcmin$ and a spatial resolution of 0.2\arcsec/pixel, and with a spectral resolving power ranging from \textit{R}$\simeq$1770 at 4800\,\AA\ to \textit{R}$\simeq$3590 at 9300\,\AA. Malin~1 was observed by MUSE on the night of the 17th April 2021 as part of ESO Program ID 105.20GH.001 (PI: Galaz), and the observations consist of  4 exposures of 1160 seconds each covering the centre and Northern part of the galaxy. The observations were carried out using the wide field mode with AO and using the extended wavelength range (WFM-AO-E).

\begin{figure}
\includegraphics[angle=0,width=1\linewidth]{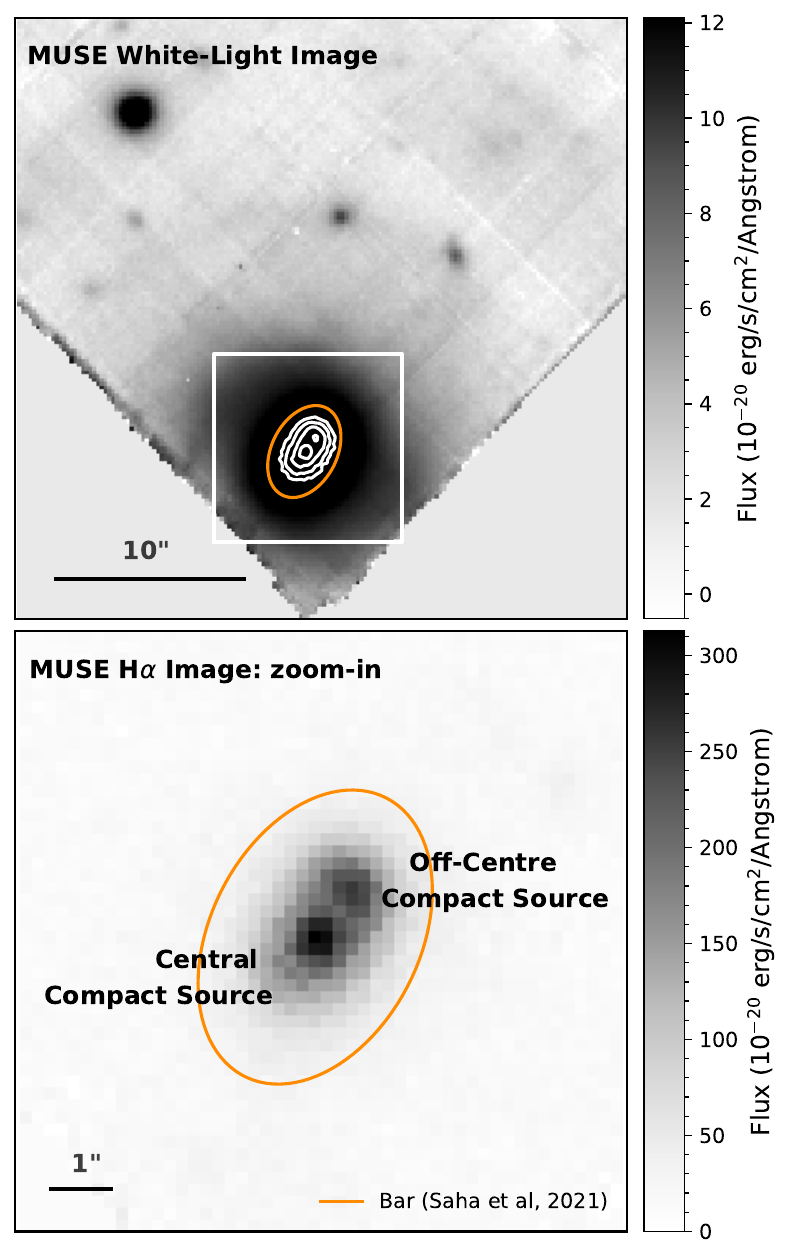}
\caption{MUSE images of Malin~1, showing the integrated white-light image on the top and the zoom in image of the nuclear region in continuum-subtracted H$\alpha$ on the bottom. The field of view in the top image is 32\arcsec\ across while the zoom-in image is 10\arcsec\ across, with that field marked with the white box in the continuum image. The contours in the top figure reflect the contours of the feature shown in the bottom image, 
and the orange ellipses in both plots represent the bar identified in \citet{Saha_2021}, with a radius of 2.6\arcsec\ \citep[$\sim$4~kpc as calculated by][]{Saha_2021} and centred on the central compact source, which is assumed throughout this paper to be the centre of the galaxy based on the peak luminosity. 
The central and off-centre compact sources have been labelled in the zoom-in image, and both images are orientated with North up and East to the left. 
\label{fig:MUSE_images}}
\end{figure}

The data were reduced using the ESO MUSE pipeline \citep[v2.6,][]{Weilbacher_2020} in the ESO Recipe Execution Tool (EsoRex) environment \citep{ESOREX}. The associated raw calibrations were used to create master bias, flat field and wavelength calibrations for each CCD, which were applied to the raw science and standard-star observations as part of the pre-processing steps. The standard star data was then used to flux calibrate the science data, and the sky continuum was measured using separate sky exposures and subtracted from the science data. The reduced data for each exposure were then stacked to produce the datacube. As a final step, any residual sky contamination was removed using the Zurich Atmosphere Purge code \citep[ZAP, ][]{Soto_2016}, which measures the sky level in the sky exposures that were processed in the same way as the science data.  The white-light image of the reduced datacube is shown in  Fig.~\ref{fig:MUSE_images}, and the PSF FWHM was $\sim0.8$\arcsec\ in the r-band image created using EsoRex.

\section{Modelling the core of Malin~1}\label{sec:Analysis}

Figure~\ref{fig:MUSE_images} shows the white-light image of Malin~1 from the MUSE datacube alongside a zoom-in of the central region of the galaxy in H$\alpha$, showing that the nuclear region is not a point source, but is instead either an extended source or  two compact sources.  The orange ellipses in Fig.~\ref{fig:MUSE_images} give the position of the bar component detected by \citet{Saha_2021}, which has a radius of 2.6\arcsec, a S\'ersic index of 0.2 and a position angle of 63\textdegree\ as measured anticlockwise from horizontal. These ellipses have been centred on the photometric centre of the galaxy, which coincides with the centre of the brighter of the two clumps. The white contours in the upper plot represent the contours of the structure seen in the H$\alpha$ image below. One can see that the contours  within the nuclear region coincide with those of the bar. Based on these contours we have labelled the two clumps as the central and off-centre compact sources (CCS and OCCS, respectively).

This section will map out the inner region of Malin~1 to investigate the nature of this structure, and aims to determine whether it is a nuclear disc, a bar or globular clusters accreting into the centre of the galaxy.

\subsection{Modelling the kinematics at the core of Malin~1}\label{sec:kinematics}
\citet{Saha_2021} found that the centre of Malin~1 contained an extended source that they considered to be a bar containing ongoing, asymmetric star formation, leading to a clumpy appearance. Other scenarios that could explain an extended compact source at the centre of a galaxy could be an eccentric stellar disc that formed from the disruption of an infalling GC \citep{Tremaine_1995}, where the eccentricity of the disc leads to the appearance of two bright sources. Evidence of a bar or a nuclear disc would be seen in the 2D kinematics in that region: a nuclear disc would present a clear rotation distinct from the underlying disc and an asymmetric velocity dispersion profile within the nuclear disc, while a bar would display a region in which the stellar velocity and $h_3$ are correlated \citep{Bureau_2005,Iannuzzi_2015}.

As a first step towards modelling the kinematics in this region, the data was spatially binned  using the Voronoi tessellation method of \citet{Cappellari_2003}. Following \citet{Junais_2023} we used a target SNR of 20 and only included spaxels with a minimum SNR of 2 in the range $6050-6500$\AA. The stellar kinematics were then modelled using the penalized Pixel Fitting (\textsc{pPXF}) code of \citet{Cappellari_2004} and \citet{Cappellari_2017}, using both stellar template spectra of well defined ages and metallicities and emission line spectra to represent the emission features. The stellar template spectra were taken from the MILES stellar library \citet{Sanchez_2006}, and both the stellar and gas template spectra were convolved with line-of-sight velocity distributions and velocity dispersions to obtain a model spectrum that best fits each binned galaxy spectrum, and low-order additive polynomials were applied to model the flux calibration of the continuum and reduce template mismatch.

\begin{figure}
\includegraphics[angle=0,width=1\linewidth]{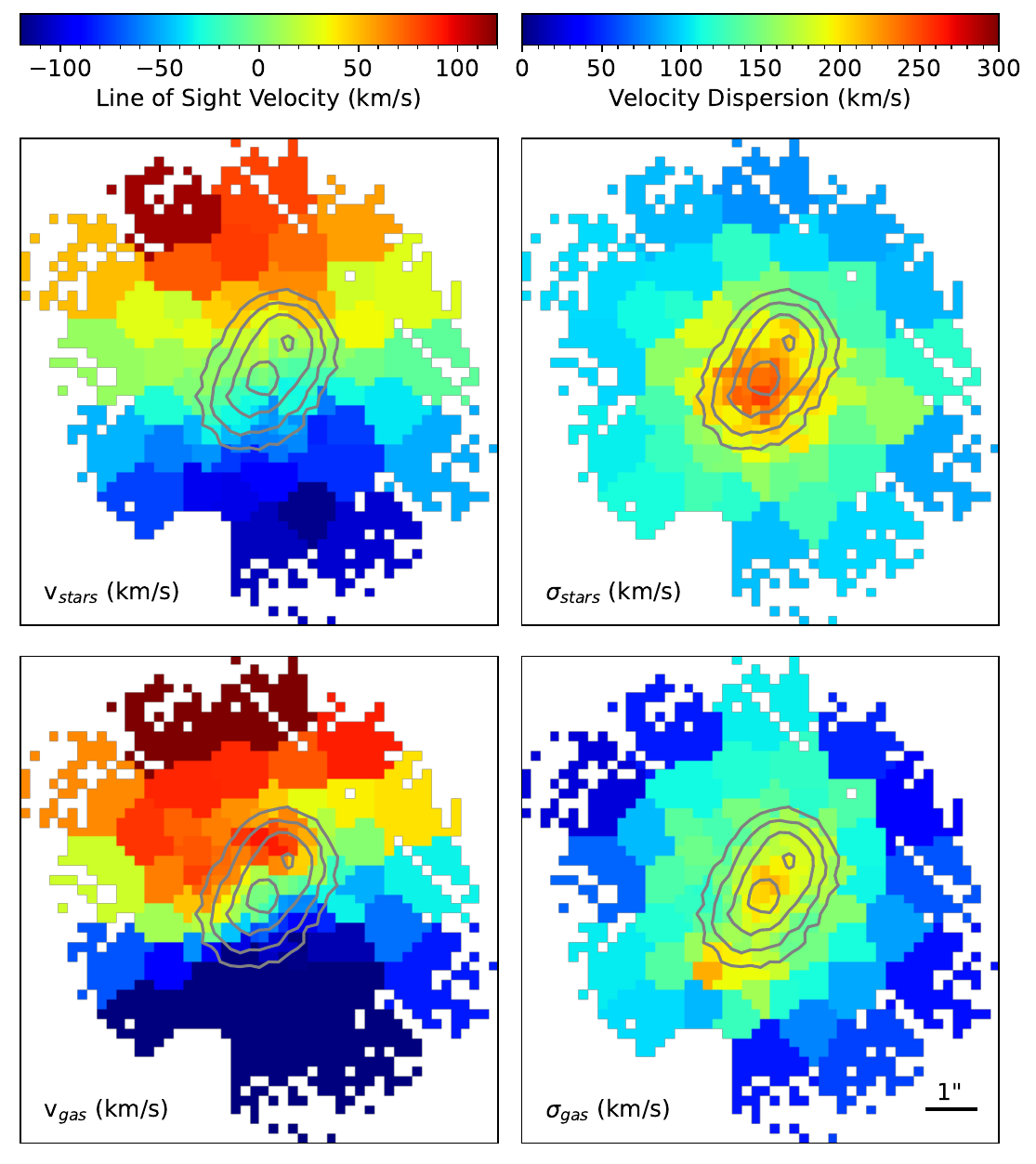}
\caption{ Maps of the stellar (top) and gas (bottom) velocity (left) and velocity dispersion (right) in the central region of Malin~1. The field of view is the same as the bottom plot in Fig.~\ref{fig:MUSE_images}, and the contours from that figure are recreated here for reference. A scale bar is given in the bottom right of the gas velocity dispersion plot.  
\label{fig:MUSE_kinematics}}
\end{figure}

\begin{figure}
\includegraphics[angle=0,width=1\linewidth]{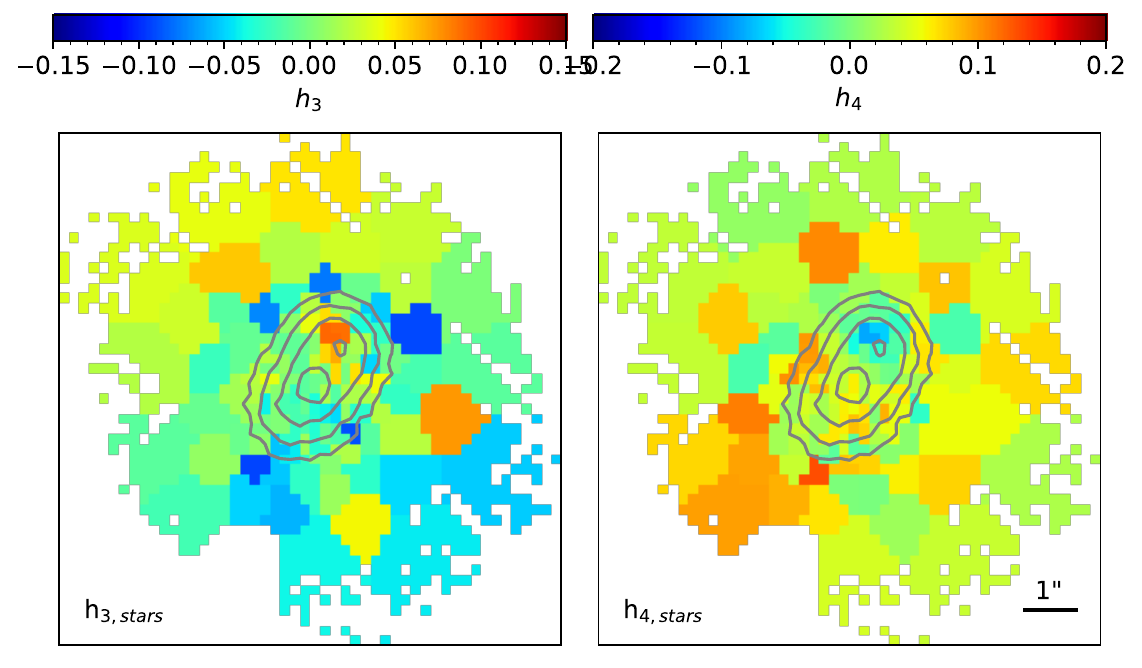}
\caption{ Same as Fig.~\ref{fig:MUSE_kinematics} but showing the stellar $h_3$ (left) and $h_4$ (right)  in the central region of Malin~1. } 
\label{fig:MUSE_kinematics_h3_h4}
\end{figure}

The stellar and gas kinematics maps can be seen in Fig.~\ref{fig:MUSE_kinematics} and the stellar $h_3$ and $h_4$ maps are plotted in Fig.~\ref{fig:MUSE_kinematics_h3_h4}, with the contours of the nuclear region from Fig.~\ref{fig:MUSE_images} overplotted for reference. No clear distortion is seen in the velocity and velocity dispersion maps within the region of the contours, indicating that the kinematics in this region are dominated by the underlying disc. The $h_3$ and $h_4$ maps, however, so appear to show different values in the areas around the two compact sources, which could indicate the presence of a distinct kinematic component. The nature of this component will be explored in more detail in Section~\ref{sec:nature}.

\subsection{Modelling Malin~1 with \textsc{buddi}}\label{sec:Buddi}

Another technique to study the central region of Malin~1 is to model the surface brightness profile to derive the properties of each component present. Furthermore, by modelling and subtracting the brighter components present, fainter and more asymmetric structures may appear in the residual image that would normally be lost in the original data where the other components dominate the light. Thus, this method could be used to study the central region of Malin~1 by subtracting the light from the rest of the galaxy and providing a better view of the nuclear structure. 

While this method has been widely applied to photometric data, only in the last decade have IFU data had sufficiently high spatial resolution and wide field of view to apply the same techniques. One code that can apply this modelling to an IFU datacube to extract the clean spectra of each component included in the fit is \textsc{buddi} \citep[Bulge--Disc Decomposition of IFU data;][]{Johnston_2017}.
\textsc{buddi} is a wrapper for \textsc{GalfitM} \citep{Haeussler_2013}, which is a modified form of \textsc{Galfit} \citep{Peng_2002,Peng_2010} that can model multi-waveband images of a galaxy simultaneously, thus using information from the entire dataset to fit each image by using user-defined Chebyshev polynomials to constrain the variation in the structural parameters over the full wavelength range. As a consequence, \textsc{GalfitM} can derive reasonable estimates of the structural parameters even for images with lower S/N where \textsc{Galfit} would fail.

\textsc{buddi} takes this idea one step further by modelling the galaxy in every image slice within an IFU datacube. Due to the large number of image slices in an IFU datacube, applying \textsc{GalfitM} directly to the datacube is impractical in terms of computing power and time required. \textsc{buddi} therefore overcomes this issue by binning the datacube and carrying out the fits in a systematic and reproducible way that can be completed much more rapidly. The final products created by \textsc{buddi} are wavelength-dependent models of each morphological component included in the fit, from which the spectra representing the light purely from that component can be extracted either as a 1-dimensional spectrum or in the form of a datacube. These spectra are consequently free from  contamination from neighbouring light sources, thus allowing an independent analysis of each component included in the model.

For this study of Malin~1, \textsc{buddi} is the perfect tool to isolate and extract the spectra of the two compact sources at the centre of Malin~1. By modelling the surface brightness profile of the rest of the galaxy, we can disentangle the spectra of each component included in the model with minimal contamination from the superposition of light from other structures. Thus, with the clean spectra extracted for each compact source in the centre of Malin~1, we can study their properties for the first time. Full details of how \textsc{buddi} works can be found in \citet{Johnston_2017}, and an overview of modelling dwarf galaxies and their NSCs in the Fornax Cluster with \textsc{buddi} are described in \citet{Johnston_2020}. However, for completeness, a brief overview of how \textsc{buddi} was used to model the galaxy and extract the spectra of the compact sources is given in Sections~\ref{sec:step0} to \ref{sec:step4} below.

%
%
%
%

\subsubsection{Step 0: Create a PSF datacube and bad pixel mask}\label{sec:step0}
Before running the datacube through \textsc{buddi}, a Point Spread Function (PSF) datacube and bad pixel mask must first be created.  The PSF datacube models how the PSF varies as a function of wavelength throughout the datacube, thus modelling the smearing of the image due to turbulence in the atmosphere. To create the PSF datacube, two stars were identified within the MUSE datacube FOV. At each wavelength within the datacube, postage stamp images of these two stars were extracted and stacked, and modelled with a Gaussian profile to create a PSF model at each wavelength. Combining these model images thus produced the PSF datacube.

Another important preparation is the bad pixel mask. It can be seen in Fig.~\ref{fig:MUSE_images} that the square MUSE FOV lies at an angle within the image, resulting in many spaxels outside of the MUSE FOV that contain no light and thus show a flux value of 0. 
Therefore, a badpixel mask was created to mark those spaxels that fall outside of the MUSE FOV and thus eliminate them from the fit.

As can be seen in Fig.~\ref{fig:MUSE_images}, the FOV also contains several bright foreground and background objects. Therefore, to minimize their impact on the fit, the regions around these objects were also included in the badpixel mask. A second badpixel mask was created that also masked out the regions of bright emission within the spiral arms. This mask allowed for a better fit to the underlying stellar disc with minimal contamination from the emission regions. Finally, a third bad pixel mask was created that also masked out the emission region around the off-centre compact source but left the central compact source unmasked. For this latter mask, we used a circular aperture centred on the OCCS in the residual datacube from an earlier fit to determine which pixels to mask out. The aim of this mask was to reduce the effect of the off-centre compact source in the fit to the central regions of the galaxy, thus helping to confirm whether the off-centre compact source is a distinct object. The steps outlined in the next sections were carried out using all three of these masks, and in in each of these fits the models for each component within the galaxy were similar, and the emission line analyses showed consistent results for the CCS and the off-centre compact source for each fit.  For simplicity, throughout the remainder of this paper, we present only the results using the third bad pixel mask that masks out the emission line regions in the spiral arms and the OCCS from the fit.

\subsubsection{Step 1: Obliterate the kinematics}\label{sec:step1}

When taking an image slice of a rotating galaxy at a wavelength close to a strong spectral feature, the galaxy may appear asymmetric. One side may be brighter/fainter than the other since it displays light in the stellar continuum while the other side is red or blueshifted into an absorption/emission feature, with further distortion due to the variation in the velocity dispersion as a function of radius within the galaxy. Since \textsc{GalfitM} cannot model non-axisymmetric structures, these effects must be normalised. Consequently, the first step carried out by \textsc{buddi} is to measure and obliterate the stellar kinematics across the galaxy, thus ensuring that each image slice within the datacube shows an image of the galaxy at the same rest-frame wavelength.

As in Section~\ref{sec:kinematics}, this step is carried out by first binning the datacube using the Voronoi tessellation technique, and then measuring the kinematics of the binned spectra with \textsc{pPXF}. 
Once the kinematics across the galaxy were measured, the spectra in each spaxel within the datacube were shifted to the closest spectral pixel to correct for the line-of-sight velocity, and broadened to match the maximum velocity dispersion measured within the galaxy. In order to reduce the effects of erroneous measurements due to the background noise, a limit of S/N~=~3 per pixel at $\sim6000$\AA~was implemented for the binning step to only include those spaxels that were not dominated by noise.

It should be noted that this correction only applies to the stellar kinematics within the galaxy. When the galaxy also contains gas that corotates with the stars with similar velocities, the kinematics corrections will have the same effect on the gas emission lines, normalizing their line-of-sight velocities and broadening their line profiles across the datacube. If the gas corotates with the stars with a difference in the line-of-sight velocity (i.e. $>1$~pixel), the kinematics corrections will either under or over correct the gas kinematics. The result of this effect is that in the final spectra extracted for each component, the gas emission lines will be broader than in the original datacube, but no flux should be lost since the spectrum is integrated from across the whole galaxy. This latter scenario was found to be true for Malin~1 where the gas was found to corotate with the stars with only a small difference in velocity outside of the central regions of the galaxy. A deeper analysis of the gas kinematics will be carried out in a future work, and an analysis of the gas properties across the galaxy within this datacube is presented in \citet{Junais_2023}.

\subsubsection{Step 2: Model the white-light image}\label{sec:step2}
Once the kinematics have been obliterated, the light profile fits can begin. The first step in the fitting process is to create the white-light image of the galaxy, and to fit that image to determine the best-fitting model. This image represents the maximum S/N image of the galaxy, taking advantage of the information over the full spectral range. Thus, this step provides a quick way to determine the best initial estimates for the physical parameters (integrated magnitude, $m_{tot}$; effective radius, $R_e$; Sérsic index, $n$; axis ratio, $q$; and position angle,   $PA$), and number of components for the fits.

\begin{figure*}
\includegraphics[angle=0,width=1\linewidth]{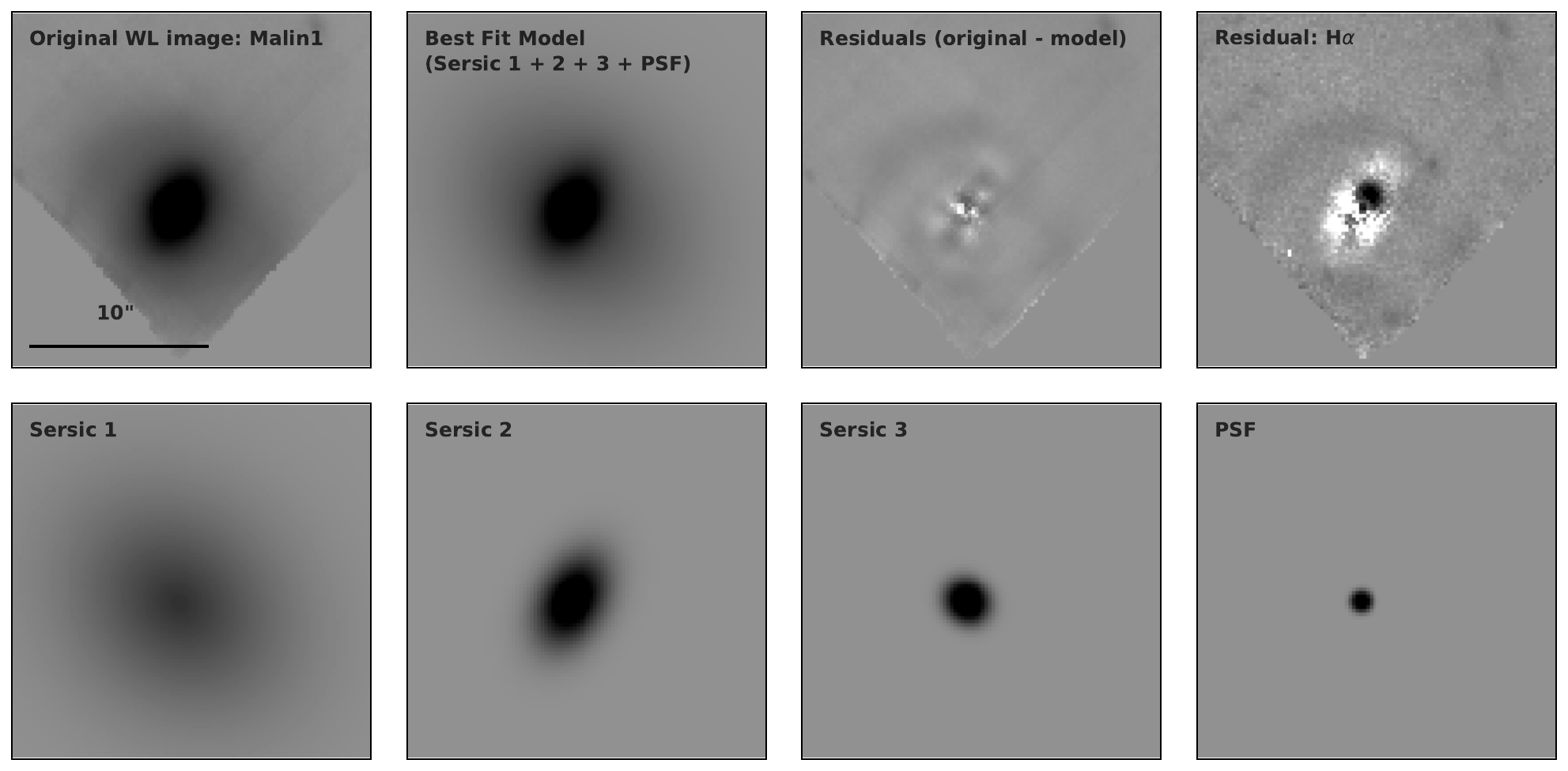}
\caption{An example of the fit to Malin~1, zoomed in to show a FOV of $20\times20\arcsec$.
 From left to right, the top row shows the white-light images from the original datacube, the best fit model, the residual image from this fit, and the residual image in the region of the H$\alpha$ wavelength, while the bottom row shows the three S\'ersic components that model the host galaxy and the PSF component that models the brighter source at the core, i.e. the CCS. All images are oriented with north up and east to the left, and use the same flux scaling. A scale bar indicating 10\arcsec\ is given in the top-left plot. 
 \label{fig:fit_overview}}
\end{figure*}

In this case, a good model for the host galaxy was achieved using three S\'ersic components, with a PSF component representing the  central compact source, as shown in Fig.~\ref{fig:fit_overview}. No constraints were used within the fits to force all four components to be centred on the same pixel, and in the final fit the centres of all components were found to be within $\sim0.5$~pixels ($0.1\arcsec$) of each other. While a discussion on the nature of the three S\'ersic components used to model the galaxy is beyond the scope of this paper, the three component fit is consistent with the fits to the disc, bulge and bar components within this galaxy by \citet{Barth_2007}, \citet{Junais_2020} and \citet{Saha_2021}. We attempted to include a second PSF component to model the off-centre compact source in the nuclear region shown in Fig.~\ref{fig:MUSE_images}, but this source was found to be too faint in the continuum to achieve a good model, and the fits failed to converge. 


\subsubsection{Step 3: Model the narrow-band images}\label{sec:step3}
Once a good fit has been achieved to the white-light image, the wavelength information can be introduced. The datacube was rebinned into a series of 10 narrow-band images along the wavelength direction, where these images have higher S/N than any individual image slice within the datacube. These images were created by stacking equal numbers of image slices along the datacube with wavebands ranging from $\sim350$\AA\ in the blue to $\sim600$\AA\ in the red due to the logarithmic binning of the datacube in the wavelength direction. The narrow-band images were then modelled, using the results from the previous step as the initial estimates. The fit was run twice, first with a polynomial of order 10 to allow full freedom in the fit parameters, simulating the effect of running \textsc{Galfit} on each image independently. This step is useful to get an idea of how the structural parameters vary as a function of wavelength, and in this case to understand the effect of emission in these narrow-band images on the fits to the galaxy. However, each narrow-band image was created by combining $\sim380$ image slices with only a small fraction of these slices showing emission lines, and upon inspection there was found to be little effect in the narrow-band images that cover a strong emission line. Furthermore, the free fits to these images also showed no obvious variations in the fit parameters between the fits to images containing emission and those displaying only stellar continuum. 

The second fit then imposes restrictions in the structural parameters by using polynomials of order 2 (i.e. linear variation with wavelength) for the $R_e$, $n$, $PA$ and $q$ while still allowing $m_{tot}$ to have full freedom. It should be noted that this order is how \textsc{GalfitM} interprets the order of the polynomial, where order 0 means the parameter is not allowed to vary from the initial estimate, order 1 is allowed to vary but must remain constant with wavelength (Chebychev order 0) and order 2 indicates a linear variation with wavelength (Chebychev order 1). Thus, information on the variation in the structural parameters as a function of wavelength was derived. It was found that, in general, the structural parameters for components all three S\'ersic components were relatively constant with wavelength, with the scatter being $<10\%$ of the mean value in all parameters, and in many cases of the order $2-3\%$.

\subsubsection{Step 4: Fit the image slices and extract the spectra}\label{sec:step4}
The next stage used the polynomial fits from the previous step to fit the individual image slices in the kinematics-obliterated datacube.  The datacube was first split up into batches of 10 consecutive image slices to reduce the computing time and avoid memory issues, and these images were then modelled by \textsc{GalfitM}. For these fits, the structural parameters ($R_e$, $n$, $q$ and $PA$) were held fixed for each image slice according to the values that were determined by the polynomial fits in the previous step, while $m_{tot}$ was left completely free. It is possible that the structural parameters for the S\'ersic components can be quite different in image slices corresponding to the wavelengths of emission features compared to the stellar continuum \citep[see e.g.][]{Johnston_2012}. However, in this study we are interested in modelling the underlying stellar continuum only and have masked out the emission line regions in the image slices at those wavelengths. Therefore, the effect on the fit due to any faint emission present in the unmasked regions is considered to be negligible.

\begin{figure*}
\includegraphics[angle=0,width=1\linewidth]{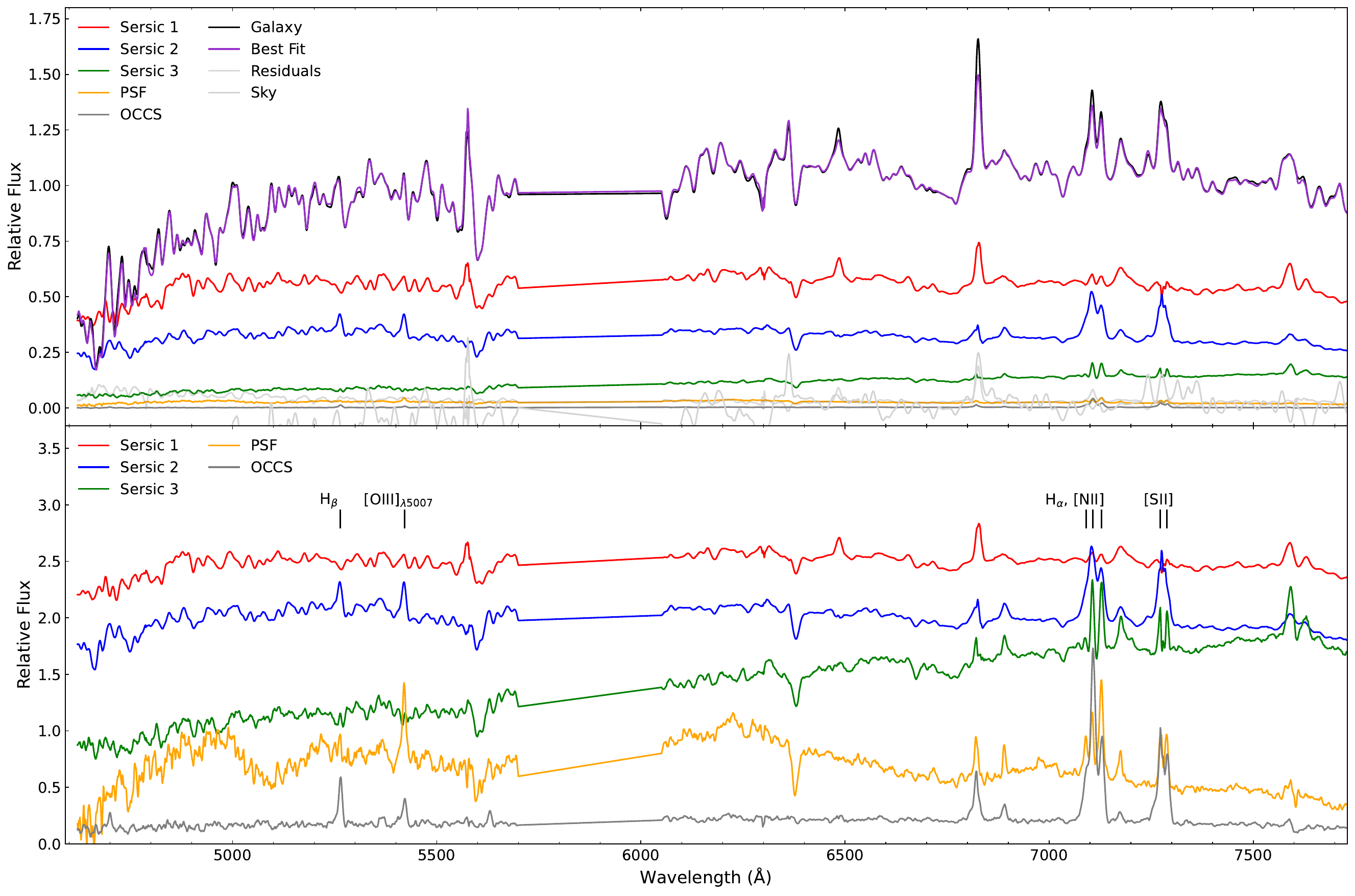}
\caption{An overview of the spectra extracted for each component shown in Fig.~\ref{fig:fit_overview} by \textsc{buddi} and the off-centre compact source (labelled as OCCS). The integrated spectrum from the original datacube is given in black with the spectrum of the best fit given in purple. The spectra for the three S\'ersic components, the PSF component representing the CCS and the off-centre compact source are shown in red, blue, green, orange and dark grey, respectively, and the residuals are given in lighter grey. The top plot shows the spectra relative to the flux of the original datacube, while the bottom plot normalises the spectra for the four components to the flux value interpolated from the middle of the notch region and offsets them arbitrarily to better display the differences in gradients and line strengths. Note that in the bottom plot, to better compare the flux of the off-centre compact source it was normalised to the flux of the PSF component in the notch region. The strong feature at 5577\AA\ is a sky line.}
\label{fig:spectra}
\end{figure*}

Once these fits were completed, the spectrum for each component was created both as a one-dimensional integrated spectrum for that component and as a model datacube. The one-dimensional spectra were derived by plotting the integrated flux from \textsc{GalfitM} for each component as a function of wavelength, and an overview of these 1D spectra are given in Fig.~\ref{fig:spectra}. The datacubes were created by stacking the  best-fitting model images of each component at each wavelength from \textsc{GalfitM}. A residual datacube was also created by subtracting the model datacubes for each component from the kinematics-obliterated datacube. A zoom in of the core region in this residual datacube can be seen in Fig.~\ref{fig:fit_overview} in the stellar continuum and in H$\alpha$ emission. There are few artefacts in this region in the stellar continuum image, such as a faint spiral arm, but the off-centre compact source becomes clear in the continuum-subtracted H$\alpha$ emission image. Similarly, there appears to be some over-subtraction of the flux in the H$\alpha$ emission image, especially around the region of the PSF component representing the CCS. The flux ratio of the pixel with the highest oversubtraction relative to the peak H$\alpha$\ emission in the area of the off-centre compact source is of the order 0.14. This artefact is likely to have several causes. For example, the PSF model used to model the CCS may not have been perfect for the MUSE data at the position of the centre of the galaxy. As described in Section~\ref{sec:step0}, the PSF datacube was created using the variations in the 2D structure of 2 stars within the MUSE FOV. Ideally more stars would have been used, but in this datacube there were no other suitable candidates (i.e. confirmed bright stars based on their 2D light profile and spectra that are far enough from the edge of the FOV and without any bright structures from the galaxy nearby). Additionally, the residual structures of the slicers and channels in the final MUSE datacube makes it hard to create a perfect PSF for the exact position that it will be used in the model. And finally, it is possible that if the CCS is a distinct source, it may not be perfectly modelled as a PSF component, especially in the emission. Since that structure was being modelled across the entire datacube, the PSF component was found to be the best fit (compared to a S\'ersic model for example) over the majority of the wavelength range in the stellar continuum. However if it is more extended in the H$\alpha$ emission, then the excess emission that falls outside of the PSF profile might instead be included in the model for another component, such as S\'ersic~2 in this case, leading to an oversubtraction of the H$\alpha$\ emission in that region. Ideally one would want to modify the models used over the wavelength regions of emission lines to account for the different physical distribution of the stars and gas, but currently that step is not within the capabilities of \textsc{buddi}.

Consequently, while the continuum for the off-centre compact source couldn't be modelled with \textsc{buddi}, its emission spectrum, with minimal contamination from the light of the rest of the galaxy and the CCS, could still be extracted from this residual datacube using a circular aperture of 4~pixels ($0.8\arcsec$)  in radius centred on the source. This aperture was selected since it covered the brightest region of the emission line component without including any light from the regions where the H$\alpha$ emission may have been oversubtracted. As a test, the analysis carried out in Section~\ref{sec:analysis} was repeated with apertures of 3 and 5 pixels ($0.6-1.0\arcsec$) in radius, with no significant difference in the results.

Figure~\ref{fig:spectra} displays the extracted spectra for all components.  The top plot gives the spectra normalised relative to the integrated flux of the galaxy, where the integrated spectrum from the original datacube (shown in black) was created by masking out the foreground and background objects in order to reduce their contamination in the spectrum. It should be remembered that these spectra represent the total integrated light for each component integrated from their centre out to infinity. Similarly, some emission lines can be seen in these spectra despite our steps to mask out the emission lines in the fits. These lines represent faint emission lines present throughout large regions of the galaxy that are generally too faint to measure in the spectra from individual spaxels or from smaller binned regions, and which again appear stronger in these spectra due to the large areas of the FOV over which the fluxes have been integrated. The bottom plot shows the spectra of each component normalized relative the to mean of their flux differences either side of the region masked out by the notch filter, i.e. $\sim5870$\AA, and offset by an arbitrary amount to allow a better comparison of their colour gradients and spectra. Since the continuum flux for the spectrum extracted for the off-centre compact source is very faint compared to the emission lines, we instead normalised this spectrum using the mean flux of the CCS component and added an offset to the flux to align it with the other spectra in this figure for easier comparison of the spectrum shape.


\section{Analysis}\label{sec:analysis}
\subsection{The structure at the centre of Malin~1}\label{sec:nature}
The central region of Malin~1 has a complex morphology, and the aim of this paper is to investigate the different possibilities and determine whether it hosts a nuclear disc, a star-forming bar, or a GC falling into an NSC. The kinematics maps in in Figures~\ref{fig:MUSE_kinematics} and \ref{fig:MUSE_kinematics_h3_h4} can be used to look for signatures of kinematic components distinct from the main disc, such as a nuclear disc or a bar. 

We will first consider the possibility of a nuclear disc. A nuclear disc would appear in the velocity maps as a region with clear rotation that is distinct from the underlying disc. In the velocity plots in Fig.~\ref{fig:MUSE_kinematics} there is no clear indicator of a nuclear disc in the center of the galaxy, with relatively symmetric velocity curves seen in both the stellar and gas maps. Similarly, both the velocity dispersion maps also appear relatively symmetric. However, the measurements in Fig.~\ref{fig:MUSE_kinematics} mainly show the dominant velocity and velocity dispersion present, while the $h_3$ and $h_4$ polynomials could provide evidence of overlapping components with distinct kinematics. \citet{Cole_2014} and \citet{Gadotti_2020} describe the signatures for a nuclear disc that formed through dissipative processes (e.g. through bar driven secular evolution) as elevated values of $h_4$ and elevated absolute values of $h_3$. These parameters from the stellar kinematics measurements are plotted in Fig.~\ref{fig:MUSE_kinematics_h3_h4}, and appear to show variations in both parameters within the contours marked on the maps.  In particular, the $h_3$ and $h_4$ values for the two brightest regions marked by the contours are different, with the upper right region, labelled the OCCS in Fig.~\ref{fig:MUSE_images}, displaying more positive $h_3$ and more negative $h_4$ than the lower left region (the CCS), and the lower left region values are more consistent with those or the surrounding disc. This result appears to indicate that the region marked by the contours is unlikely to be a nuclear disc with distinct kinematics relative to the underlying disc, and instead may host two structures with distinct kinematics.

We will now consider the possibility of a bar. \citet{Bureau_2005} and \citet{Iannuzzi_2015} describe the kinematic signature of a bar as being a region in which the stellar velocity and $h_3$ are correlated due to the superposition of a bar and a disc contributing similar fractions of the light. Such signatures are presented for MUSE observations of barred galaxies in \citet{Gadotti_2020}, showing that as you move from the centre of the galaxy outwards the stellar velocity and $h_3$ are anticorrelated, then correlated, and then anticorrelated again. The maps in Figures~\ref{fig:MUSE_kinematics} and \ref{fig:MUSE_kinematics_h3_h4} are harder to interpret since the potential bar (identified by the contours) and the stellar disc kinematics are not aligned. While there might be a correlation within the contours, such that the upper right of the contoured region has higher $h_3$ and higher velocity than the lower left region, the amplitudes of $h_3$ in these regions are asymmetric, making it difficult to be sure of this correlation. Another possibility is that the OCCS has been captured by the bar and it is orbiting in an x1 orbit going along the bar while corrotating with it. This would naturally explain why both have almost the same alignment, while in the eccentric disk scenario in principle the alignment would be by coincidence. Therefore, from the kinematics it is unclear whether this region is a bar, as proposed by \citet{Saha_2021}.

If this region does contain a bar, the clumpy appearance could be due to asymmetric star formation. Such a scenario has been seen in the lenticular galaxies IC~676 \citep{Zhou_2020} and PGC~34107 \citep{Lu_2022}, and in a number of lenticular and spiral galaxies studied by \citet{Diaz_2020}, where these galaxies were found to host large-scale bars containing asymmetric star formation in young massive clusters. The origin of the gas fuelling the star formation in those clusters could come from two sources- external gas accreted through a merger \citep{Kruijssen_2012}, or disc gas losing its angular momentum in the presence of a bar, leading it to migrate in to the inner regions of the galaxy where it can trigger a new episode of star formation \citep{Athanassoula_2013,Consolandi_2017}. One can use the angular momentum vectors of the stars and gas to distinguish between these two scenarios, where the accretion of gas scenario would result in the gas kinematics being misaligned relative to the stellar kinematics, while the bar-driven inflow scenario would give more aligned kinematics. One can see in Fig.~\ref{fig:MUSE_kinematics} that the stellar and gas kinematics are relatively well aligned throughout the bar and disc. The global kinematic position angles for the stars and gas velocities were measured using the \textsc{PaFit} package, which implements the algorithm outlined in Appendix~C of \citet{Krajnovic_2006}. The kinematic position angles for the stars and gas were found to be $9\pm9$\textdegree and $17\pm3$\textdegree as measured anticlockwise from the vertical, respectively. Both these position angles are consistent within their uncertainties, indicating that the gas fuelling the star formation in the point sources at the centre of Malin~1 is most likely to originate from gas migrating in from the extended disc, possibly driven by a bar if present.

While the evidence for a nuclear disc or bar in the inner region of Malin~1 is not very strong, the kinematics maps do appear to show that the region around the OCCS has distinct kinematics relative to the CCS and outside of the contours. Thus is appears likely that the OCCS is a kinematically distinct component, which may be a star cluster falling into the centre of the galaxy or a star forming region within a bar if it's present.

The next step in the analysis was to model the light profile of the galaxy with \textsc{buddi}. Images of each component included in the fit are given in Fig.~\ref{fig:fit_overview} and show that the component labelled S\'ersic~2 does appear to have a bar-like structure with a similar size and position angle to the bar modelled by \citet{Saha_2021} (this work: $R_e=1.33\arcsec, n=0.99, PA=64$\textdegree. \citet{Saha_2021}: $R_e=1.67\arcsec, n=0.20, PA=63$\textdegree). 

The fits to the light profile required a PSF component to model the very centre of the galaxy, the region identified as the CCS, but was unable to model the OCCS since it was too faint in the continuum. Instead the spectrum of that component was extracted from the residual datacube. One must always be careful when modelling the light profiles of galaxies to ensure that the components included in the model are real and thus represent a true physical structure within the galaxy. For example, a study of the light profile fits to Elliptical galaxies by \citet{Huang_2013} found that while the majority of their sample could be well modelled with three S\'ersic components, around 20\% of their sample required an additional components. These components modelled either real structures present within the galaxy, such as edge on discs or bars in the case of misclassified S0s, or were used to compensate for irregular structures such as dust lates that affected the light profile. Therefore in the following subsections we will investigate in more details the nature of the CCS and OCCS from their extracted spectra.

\subsection{Confirmation of the location of the CCS and OCCS}\label{sec:location}

If we assume that the CCS and OCCS are distinct sources, one scenario could be that they represent a star cluster falling into an NSC. According to \citet{Boker_2002}, NSCs are dense but resolved star clusters that are often the brightest and only cluster within a kiloparsec from the photometric centre of the galaxy, and \citet{Neumayer_2011} adds to that by showing that NSCs occupy the kinematic core of their host galaxies. The light profile fits in Section~\ref{sec:step2} found that the centre of the CCS, which was modelled by the PSF component, lies within 0.5~pixels (0.1\arcsec) of the centre all the other components included in the fit, indicating that it lies at the photometric centre of the galaxy. However, the fact that it could only be modelled as a PSF profile indicates that the region covered by the CCS is not resolved, making it difficult to determine if it is an NSC or whether it contains light from multiple nuclear components. The PSF FWHM was $\sim0.8\arcsec$, which corresponds to a distance of 1.4~kpc. Within this region the stellar absorption could originate from a nuclear bulge or disc or a star cluster, while the emission lines could be emitted from a nuclear ring.

In order to better understand the nature of the CCS, the kinematics were measured from the spectrum extracted for the PSF component using \textsc{ppxf} following the same methodology as described in Section~\ref{sec:step1} except that a gas component was included in the fit.  The stellar and gas kinematics for this component are consistent within $1\sigma$ ($v_{LOS,stars}=23771\pm16$km/s, $v_{LOS,gas}=23802\pm15$km/s), indicating that the origin of the stellar absorption and emission lines within the CCS spectrum lie at the same position in the galaxy. The stellar and gas velocities of the binned spectrum from the original datacube in the region of the CCS are $v_{LOS,stars}=23818\pm18$km/s and $v_{LOS,gas}=23816\pm16$km/s with velocity dispersions of $\sigma_{stars}=253\pm12$~km/s and $\sigma_{gas}=192\pm10$~km/s, which indicate that the kinematics of the CCS component are consistent with the kinematic core of the galaxy.

The kinematics of the OCCS were measured in the same way, but omitting the stellar absorption spectrum due to the low S/N. This structure was found to have a gas velocity of $v_{LOS,gas}=23850\pm16$~km/s, which is consistent with the gas velocity of the CCS to within $\sim2.1\sigma$, indicating that these two compact sources are likely to be neighbours.



Since we have the coordinates of the centres of the CCS from the model of the PSF component and the off-centre compact source from the fits with \textsc{GalfitM}, which are given in Table~\ref{tab:fluxes}, and the parameters for the 2D model fit to the disc of the galaxy (S\'ersic~1), we can calculate the deprojected distance between these two objects. The model fit to the extended disc gave 
a disc major axis with a $PA$ of 48.7\textdegree (measured anticlockwise from vertical), an axes ratio, $b/a$, of 0.82, which results in an inclination of 34.9\textdegree. This axes ratio corresponds to an ellipticity of 0.18, which is consistent with that found by \citet{Saha_2021}. With a separation on sky of 1.05\arcsec or 1.925~kpc and assuming that the sources lay at the main plane of the galaxy, we calculate that the off-centre compact source has a deprojected distance of $2.347\pm0.013$~kpc from the CCS. For comparison, using HI maps of Malin~1 \citet{Lelli_2010} measured the $PA$ to be 0 and the inclination, $i$ to be $38\pm3$\textdegree, which results in a deprojected distance of $2.45\pm0.11$~kpc between the two objects, which is consistent with our findings.

As a final step, we can determine whether the off-centre compact source is in a gravitationally bound orbit or not, and thus whether it is likely to be falling into the CCS. Using the $PA$ of the extended stellar disc component (48.7\textdegree), and extrapolating inwards the circular velocity $v_c$ modeled in \citet{Lelli_2010}, we find that the projected $v_c$ of the OCCS is $v_{c,proj}= 130$~km/s. Taking into consideration the geometrical orientation of the HI disk changes this value to 173km/s, and mass variations of a factor of two vary this result by only 30$\%$. Either way, given the observed $v_{LOS}$ difference with the CCS of ~50km/s, the projected $v_c$ place the OCCS in a bound orbit, and possibly in a radial orbit along the bar of Malin 1.
If we assume the geometry of the HI maps from \citet{Lelli_2010} instead ($PA=0$ and $i=38$) we find that the tangential velocity is $v_{c,proj}=173.72$~km/s. The discrepancy between the geometry derived form the HI and the photometry could be due to some warp or interaction that generated an offset between the stellar and gaseous disks, or it could come from the low spatial resolution HI data. However, it is interesting that both values are less than the velocity dispersion measured at the centre of Malin~1 above and greater than the velocity difference between the two objects, indicating that the off-centre compact source may be on a decaying orbit around the CCS.

\subsection{Analysis of the emission line properties}\label{sec:BPT}

\textsc{buddi} models the image of the galaxy at each wavelength, using the constraints on the structural parameters to model the stellar continuum in each image slice. While the areas with strong emission were masked out in the fits, there is likely to be some low level emission throughout the galaxy that can be detected when integrating the light over portions of the field of view, such as the components included in the model fit. Therefore, the emission lines present in the spectra for each component, which can be seen in Fig.~\ref{fig:spectra}, represent the faint emission present throughout that component.

\begin{table*}
\caption{Details of the coordinates of the centres of each component included in the fit and the off-centre compact source (OCCS), and the flux values for the main emission lines. Those marked in grey have amplitude-to-noise ratio (ANR) $<3$, and so the data points in Fig.~\ref{fig:BPT} using those fluxes are marked with open symbols. The fluxes are in units of \mbox{$10^{-17} $erg~s$^{-1} $cm$^{-2} $\AA$^{-1}$}}             
\label{tab:fluxes}      
\centering          
\begin{tabular}{l r r r r r r r r }     
\hline\hline       
Component & Ra & Dec &  $H_\beta$ & $H_\alpha$ & [OIII]$_{5007}$  & [NII]$_{6583}$ & [SII]$_{6717}$ & [SII]$_{6731}$  \\
\hline                    
   S\'ersic 1		& 12:36:59.37 & +14:19:48.71 		& \grey{$24\pm 5$}	& $67\pm 5$ 	& $83\pm 7$ 	& \grey{$18\pm 7$} 	& \grey{$5\pm 6$} 	& \grey{$18\pm 6$} 	\\
   S\'ersic 2		& 12:36:59.38 & +14:19:48.87 		& $129\pm 2$	& $236\pm 2$ 	& $143\pm 3$ 	& $187\pm 3$ 	& $157\pm 3$ 	& $111\pm 3$ 	\\
   S\'ersic 3		& 12:36:59.38 & +14:19:48.80	&\grey{$6.2\pm 1.0$}	& $35.2\pm 0.9$ 	& \grey{$1.3\pm 1.2$} 	& $36.5\pm 1.2$ 	& $14.9\pm 1.0$ 	& $12.9\pm 1.0$\\
   PSF (CCS)			& 12:36:59.38 & +14:19:48.79 	 	& $6.6\pm 0.4$	& $10.1\pm 0.4$ 	& $20.7\pm 0.5$ 	& $20.4\pm 0.5$ 	& $6.7\pm 0.5$ 	& $6.7\pm 0.5$ 	\\
   OCCS		& 12:36:59.34 & +14:19:49.68	 	& $8.7\pm 0.2$	& $30.60\pm 0.2$ 	& $5.8\pm 0.2$ 	& $20.0\pm 0.2$ 	& $1.4\pm 0.2$ 	& $10.6\pm 0.2$ 	\\
\hline                  
\end{tabular}
\end{table*}

%

\begin{figure*}
\includegraphics[angle=0,width=0.8\linewidth]{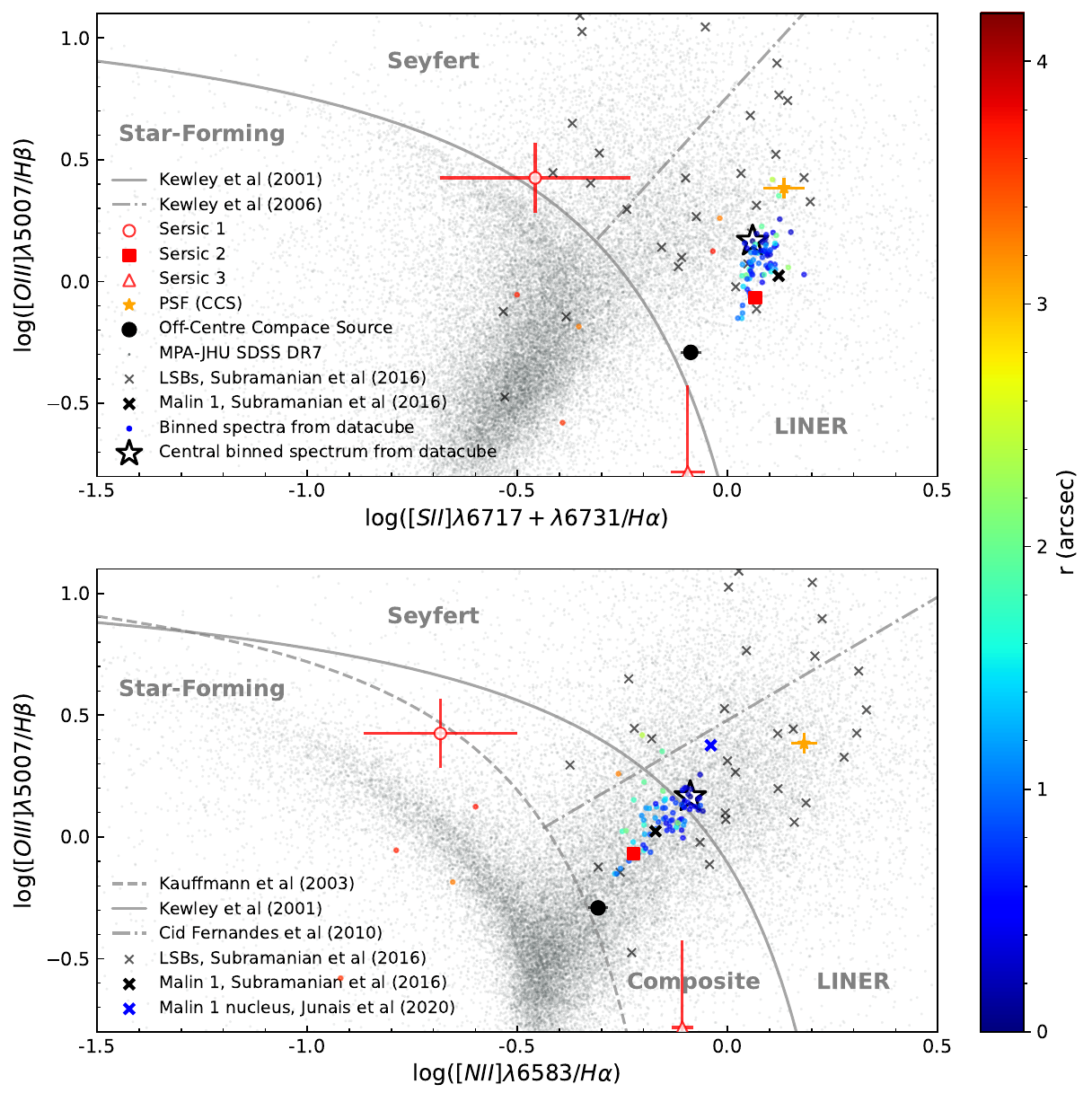}
\caption{ BPT diagrams showing the location of each component modelled by \textsc{buddi} as the larger red and yellow points, and of the off-centre compact source as the black filled circle. The open red symbols identify measurements where at least one of the emission lines used had an ANR of less than 3, and so those measurements should be used with caution. Note that the error bars for S\'ersic~2 and the off-centre compact source are within the data points. The small coloured points represent the measurements from the original galaxy datacube after binning to increase the S/N, where the colour corresponds to the distance of each binned spectrum from the centre of the galaxy according to the colour bar, and the star symbol is centred on the data point at the centre of the central PSF component. For comparison, the measurements from the MPA-JHU SDSS DR7 galaxies \citep{Brinchmann_2004} have been plotted as  grey points, and the central regions of other LSBs from \citet{Subramanian_2016} are shown as grey crosses. The black and blue crosses are the emission line ratios from the nuclear region of Malin~1 from \citet{Subramanian_2016} and \citet{Junais_2020}, respectively. The solid curves in both plots represent the maximum starburst line of \citet{Kewley_2001}, the dashed curve in the bottom plot represents the pure star-forming line of \citet{Kauffmann_2003}, and the dot-dashed lines in the top and bottom plots differentiate between the Seyfert and LINER classifications according to \citet{Kewley_2006} and \citet{CidFernandes_2010} respectively.
\label{fig:BPT}}
\end{figure*}

The BPT diagram of \citet{Baldwin_1981} is a powerful tool to compare various emission line strength ratios and use them to interpret the characteristics of the galaxy. The emission lines were measured using \textsc{ppxf}, modelling both the stellar continuum, absorption features and the gas emission lines simultaneously, using the same method described in Section~\ref{sec:location}, except that this time multiplicative Legendre polynomials were used instead of additive polynomials. While using these multiplicative polynomials slows down the fit, they have the advantage that they have less impact on the line strengths and make the fit more insensitive to dust reddening, thus eliminating the need to include a dust reddening curve. After completing the fit, the emission line strengths for the H$\alpha$, H$\beta$, [OIII]$_{\lambda 5007}$, [SII]$_{\lambda 6717}$, [SII]$_{\lambda 6731}$ and the [NII]$_{\lambda 6583}$ features were measured and plotted onto the BPT diagram shown in Fig.~\ref{fig:BPT}. The line fluxes were not corrected for extinction before plotting them onto the BPT diagram since the line rations are generally considered almost unaffected by extinction due to their proximity in wavelength \citep[see e.g.][]{Belfiore_2015}. 

In order to determine the reliability of the positions on the BPT diagram, the amplitude-to-noise ratio (ANR)  for each emission line was measured, where the noise was taken to be the standard deviation of the residual spectrum after subtracting the best fit spectrum from the input spectrum. A limit of ANR=3 was used, and where at least one emission line in a spectrum had an ANR less than this limit, the measurements were plotted as an open symbol. This situation was found to be true for components labelled as S\'ersic~1 and 3, as shown in Fig.~\ref{fig:BPT}, while S\'ersic~2 and the two compact sources were found to have all emission lines with $\text{ANR}>5$. The emission line fluxes are listed in Table~\ref{tab:fluxes} for each component included in the fit, with those lines with ANR$<3$ marked in grey.

For comparison, the process was repeated for the binned spectra extracted from the original datacube in Section~\ref{sec:step1} and these measurements are also plotted in Fig.~\ref{fig:BPT}, colour coded according to the galactocentric distance of the centre of each bin. Additionally, measurements of other emission-line galaxies from the MPA-JHU SDSS-DR7 catalog \citep{Brinchmann_2004}
have also been plotted as  small grey points for reference.

It can be seen that the most extended S\'ersic component in Malin~1, S\'ersic~1 in Figure~\ref{fig:fit_overview}, lies within the star-forming region of the BPT diagram, while the other two more compact components, S\'ersics~2 and 3, lie within the LINER or composite regions. Similarly, the line ratios for the binned spectra from the original datacube also fall within these LINER or composite regions, with a tentative gradient such that measurements from closer to the centre of the galaxy (darker blue)  lie further to the right and as one moves more towards the points further out in the galaxy, especially the few points in the orange and red colours, the points appear to lie closer to the star forming regime. The majority of the points are generally coinsistent with the position of the points for S\'ersic~2 in both plots. At this point it should be remembered that S\'ersics~1 and 3 have a lower ANR in at least one of their key diagnostic emission lines, as shown in Table~\ref{tab:fluxes}, and so these line ratios should be used with caution. 
But in general this result indicates that the ionising gas throughout the galaxy comes from a combination of star formation, AGN and shocks. These results are consistent with those of \citet{Barth_2007}, who classed Malin~1 as a LINER galaxy, \citet{Subramanian_2016}, who identified the galaxy as a LINER galaxy with weak ionisation  from both AGN and starbursts, and \citet{Junais_2020}, who found that Malin~1 lies on the border between LINERS and Seyferts on the BPT diagram. Similarly, these line ratios are consistent with those of the central regions of  other LSBs in \citet{Subramanian_2016}, which are also plotted in Fig.~\ref{fig:BPT}.

The point for the CCS  lies significantly closer to the AGN/LINER regime in both plots, and interestingly appears to be offset from the measurements from the binned spectra but following the same trend (the star symbol in the plots in Fig.~\ref{fig:BPT} mark the spectrum extracted from CCS at the centre of the galaxy). This effect has been seen before for bulges in \citet{Johnston_2017} and for NSCs in dwarf galaxies in \citet{Johnston_2020}, and simply reflects the strength of \textsc{buddi} at cleanly modelling and extracting the light from even faint components within galaxies, whose measurements would otherwise be contaminated by that from the rest of the galaxy. Similarly, the line ratios for the PSF component are offset from those from the nuclear region of Malin~1 from \citet{Subramanian_2016} and \citet{Junais_2020}, shown as the bold crosses in Fig.~\ref{fig:BPT}, which is again likely due to the clean extraction of the PSF spectrum in this work. It can be seen that the measurements from the literature are in better agreement with those from the binned spectra from the original datacube, reflecting that the spectra in the literature of the `nucleus' of Malin~1 contain contamination from the light from the rest of the galaxy. 
Thus, by using this method to cleanly extract the spectrum of the CCS we can assert that this structure is a Low-Ionization Nuclear Emission-Line Region (LINER), potentially powered by shocks, post-AGB stars and/or low-level accretion activity onto a central massive black hole. 

Finally, the OCCS appears to lie on the border with the star-forming regime in both plots. Interestingly, the line strength ratios for this object are similar to those of S\'ersic~3 in the top plot of Fig.~\ref{fig:BPT}, the most compact S\'ersic component in the fit, despite the low ANR of the emission lines in that component. However, the points for these two components do not coincide in the bottom plot, indicating that they do not experience the same mixed ionization source. 

At this point it should be noted that Fig.~\ref{fig:fit_overview} shows a possible ovsersubtraction of the H$\alpha$ emission around the OCCS, which might affect the position in the BPT diagram. As a test of this effect, we extracted two spectra of the surrounding region within annuli of 3 and 5 pixels radius centred on the 4 pixel radius of the OCCS aperture. These spectra were used to `correct' the OCCS spectrum for the possible ovsersubtraction, and the emission lines were measured and analysed in the same way as above. While the position of these corrected spectra in the BPT diagram did move slightly upwards and to the right of the position shown in Fig.~\ref{fig:BPT}, the offset was very small and within the error bars. Therefore, the results presented here appear to be minimally affected by this possible oversubtraction of H$\alpha$ and other emission lines.

Having measured the emission line fluxes for the two compact sources, it is also possible to estimate their star-formation rates (SFRs). First, the H$\alpha$ flux was corrected for dust extinction using the \citep{Calzetti_2000} extinction law with the Balmer decrement, 
\begin{equation}
F(H\alpha)_i = F(H\alpha)_o \times 10^{1.33E(B-V)},
\end{equation}
where $F(H\alpha)_o$ is the observed H$\alpha$ flux and $E(B-V)$ is the foreground dust reddening along the line of sight:
\begin{equation}
E(B-V) = 1.97 \text{log}\Bigg(\frac{F(H\alpha)_o/F(H\beta)_o}{2.86}\Bigg).
\end{equation}
We assumed an intrinsic Balmer ratio of 2.86 for case-B recombination \citep{Osterbrock_1989} at an electron temperature $T_e = 10,000$~K and density $n_e = 100 cm^{-3}$ \citep{Hummer_1987}.
The H$\alpha$ luminosity of each component, $L(H\alpha)$ was then calculated using 
 \begin{equation}
L(H\alpha) = 4 \pi d_L^2F(H\alpha)_i,
\end{equation}
where $d_L$ is the luminosity distance, and finally the SFR was determined using the relation
\begin{equation}
{\rm SFR}({H}\alpha)=1.26 \times 10^{-41} L({H}\alpha)
\end{equation}
from \citep{Kennicutt_1994}. These calculations resulted in SFRs of $0.012\pm0.005~\text{M}_\odot\text{yr}^{-1}$ for the CCS and $0.32\pm0.02~\text{M}_\odot\text{yr}^{-1}$ for the off-centre compact source. The SFRs for both compact sources are of the same order as those of the two star forming regions in the bar of the S0 galaxy IC~676, which have values of $0.08~\text{M}_\odot\text{yr}^{-1}$ and $0.19~\text{M}_\odot\text{yr}^{-1}$ \citep{Zhou_2020}. In Malin~1, the SFR of the CCS is significantly lower than the other compact source, in line with what would be expected if the ionization source is dominated by the LINER as shown in Fig.~\ref{fig:BPT}. Furthermore, in the LINER regime one should remember that the Balmer emission lines could also be excited by AGN or shocks rather than only by star formation, and so the SFR derived above from the H$_{\alpha}$ flux alone  may only be considered an upper-limit

The combination of the similar line-of-sight velocity of the OCCS and the centre of the galaxy with the emission line properties renders it possible that the off-centre compact source  consists of one or more unresolved clumps of ongoing star formation that lie very close to the centre of the galaxy, where it is too faint (i.e. low mass) to have a significant stellar continuum at the sampled wavelengths, but bright enough in emission to be detected in specific emission lines.


\section{Conclusions}\label{sec:Conclusions}
The centre of Malin~1 has long been known to be complex, with evidence of a  bar \citep[e.g.][]{Saha_2021} and emission line studies hinting at the presence of both star-formation \citep[e.g.][]{Junais_2020} and LINER \citep[e.g.][]{Barth_2007, Subramanian_2016, Junais_2020} activity. However, the small size of this bar structure \citep[semi-major axis length $\sim2.6\arcsec$ or 4.3~kpc in projection;][]{Saha_2021} has limited attempts to better understand this region. The aim of this paper is to explore the core  of Malin~1 with MUSE IFU data to determine the nature of the nuclear region. Analysis of the nuclear kinematics revealed no clear evidence of a nuclear disc or a bar with kinematics distinct from the extended disc, but did suggest that the region around the off-centre compact source contained multiple kinematic components. We then applied a careful light profile fit to the entire datacube, finding that the best model consisted of three S\'ersic components, a PSF component modelling the CCS, and finally the spectrum of the OCCS was extracted from the residual datacube. Interestingly, one of the S\'ersic components in this model is consistent with the bar identified in \citet{Saha_2021}, indicating that if the bar is real it does not show very different kinematics from the rest of the galaxy or is faint enough that it doesn't dominate the kinematics in that region. But the presence of a bar in this region is consistent with the scenario presented by \citet{Saha_2021} in which the centre of Malin~1 contains a bar hosting asymmetric star formation.

 Analysis of the emission line kinematics of the CCS and OCCS reveals that both compact sources lie at the kinematic core of Malin~1, which is consistent with the definition of an NSC by \citet{Neumayer_2011}. However, the distance to Malin~1 means that it is impossible to resolve this region and confirm that both the stellar continuum and gas emission lines in the region of the CCS have the same origin. If the CCS is in fact an NSC, this result implies that the galaxy contains an NSC onto which a lower-mass, star-forming clump is being accreted. We could, therefore, witness active NSC formation in action. The key emission line ratios were  plotted onto  BPT diagrams to explore the nature of the ionization, and reveal that the CCS is consistent with a LINER while the OCCS lies in the composite region, very close to the separation between star-forming galaxies and AGN. Additionally, the $h_3$ and $h_4$ maps show higher and lower values, respectively, in the region of the OCCS compared to the CCS and the surrounding disc, suggesting that this region contains an additional structure with distinct kinematics superimposed on the underlying disc. Thus, the large difference in the line strengths of these two targets and the trends in the  $h_3$ and $h_4$ maps further adds to the case that they are distinct objects within the core of the galaxy. Assuming that the central compact source is the original and more massive NSC since it lies closer to the kinematic and photometric centre of the galaxy (defined as the peak in the velocity dispersion and light profile, respectively) and has a brighter stellar continuum, the off-centre compact source could either be  a star-forming clump containing one or more star clusters that is in the process of falling into the core of the galaxy and which will eventually merge with the central NSC, or a clump of gas infalling into the centre of the galaxy from either outside or from the disc and triggering star formation there. These scenarios are consistent with the wet-migration \citet{Guillard_2016} and gas accretion \citet{Silk_1987} methods of assembling an NSC. However, with the resolution available with the available MUSE data we are unable to distinguish between these two models. Thus, with future deep photometric observations and additional MUSE pointings to cover the rest of the galaxy, we will be able to investigate further the nature of the core of Malin~1.



%
\begin{acknowledgements}
The authors would like to thank the referee for their useful comments which helped to significantly improve the paper.
E.J.J. acknowledges support from FONDECYT Iniciaci\'on en investigaci\'on 2020 Project 11200263. 
M.B. acknowledges FONDECYT postdoctorado 2021 No. 3210592
E.J.J., G.G., and T.H.P. gratefully acknowledge support by the ANID BASAL projects ACE210002 and FB210003.
J is grateful for support from the Polish National Science Centre via grant UMO-2018/30/E/ST9/00082.
T.H.P. acknowledges support through FONDECYT Regular Project No. 1201016.
Y.O-B acknowledges support from FONDECYT Postdoctoral 2021, project No. 3210442. 
PMW gratefully acknowledges support by the German BMBF from the ErUM program (project VLT-BlueMUSE, grant 05A20BAB).
\end{acknowledgements}

\bibliographystyle{aa} 

\end{document}